# Occulter to Earth: Prospects for studying Earth-like planets with the E-ELT and a space-based occulter

**Corresponding author:**

Markus Janson, Stockholm University, Stockholm, Sweden
Contact: markus.janson@astro.su.se

**Co-authors:**

Thomas Henning, Max Planck Institute for Astronomy, Heidelberg, Germany
Sascha P. Quanz, Swiss Federal Institute of Technology, Zurich, Switzerland
Ruben Asensio-Torres, Max Planck Institute for Astronomy, Heidelberg, Germany
Lars Buchhave, Technical University of Denmark, Kongens Lyngby, Denmark
Oliver Krause, Max Planck Institute for Astronomy, Heidelberg, Germany
Enric Palle, Instituto Astrofisica de Canarias, La Laguna, Spain
Alexis Brandeker, Stockholm University, Stockholm, Sweden

## Abstract

Direct detection and characterization of Earth-like planets around Sun-like stars is a core task for evaluating the prevalence of habitability and life in the Universe. Here, we discuss a promising option for achieving this goal, which is based on placing an occulter in orbit and having it project its shadow onto the E-ELT at the surface of Earth, thus providing a sufficient contrast for imaging and taking spectra of Earth-like planets in the habitable zones of Sun-like stars. Doing so at a sensible fuel budget will require tailored orbits, an occulter with a high area-to-mass ratio, and appropriate instrumentation at the E-ELT. In this White Paper, submitted in response to the ESA Voyage 2050 Call, we outline the fundamental aspects of the concept, and the most important technical developments that will be required to develop a full mission.

**Declarations:**
**Funding** – MJ is supported by a grant from the Knut and Alice Wallenberg (KAW) foundation.
**Conflicts of interest/Competing interests** – not applicable.
**Availability of data and material** – not applicable.
**Code availability** – not applicable.

## Exoplanet detection

The past two decades have seen an explosion in the field of exoplanet research, where the scientific consensus has gone from not knowing whether exoplanets exist altogether, to now knowing that exoplanets are nearly ubiquitous around a wide range of stellar host types (e.g. Burke et al. 2015; Dressing & Charbonneau 2015). These insights have largely been acquired through the radial velocity (RV) and transit techniques, which have provided means to efficiently detect and characterize planets in small orbits around quiescent stars (e.g. Howard et al. 2013; Coughlin et al. 2016). At increasing planet separation, however, these techniques become progressively less efficient, which in practice means that other methods are necessary to detect and characterize a wider exoplanet demographic. Methods that can provide the necessary sensitivity at wider separations (and for less quiescent stars) include gravitational microlensing, astrometry, and direct imaging. Microlensing constitutes a special case in that it has the potential to provide a statistical census of exoplanet demographics, potentially down to Ganymede-like masses (e.g. Penny et al. 2019); but the planetary systems detected with this technique will generally be so distant that they cannot feasibly be followed up past the transient lensing event, so no detailed characterization of the planets is possible via microlensing. By contrast, astrometry and direct imaging favour detections in nearby systems, which is ideal for detailed characterization.

Astrometry has so far only led to a few individual exoplanet detections (e.g. Benedict et al. 2002; Snellen & Brown 2018), but once *Gaia* releases its final catalog, this situation will drastically change, with astrometry most likely providing more exoplanet detections than RV and transits combined (e.g. Perryman et al. 2014). The astrometric signature of a star-planet system increases with increasing semi-major axis, thus resulting in an opposite bias with respect to RV and transits, and an opportunity to study the wider planet population. It is also relatively insensitive to the orientation of the target system, unlike RV, which is insensitive to near face-on systems; and even more so unlike transits, which strictly require near edge-on systems for detectability. It also yields a direct measure of the planet mass, in contrast to RV, which only provides a lower mass limit unless the system geometry can be independently constrained.

To directly image planets themselves is the most intuitively obvious way to detect them, but it requires the ability to overcome the vast brightness contrast between a planet and its parent star at a small angular separation, which until recently was an excessively difficult technical task. However, technological developments in adaptive optics and coronagraphy have been rapid, and now enable the detection of sufficiently wide, massive, and young planets with ground-based telescopes (Marois et al. 2008; Lagrange et al. 2009). Planets are easier to image (in the infrared) when they are young, since they start out with a lot of heat from their formation, and then gradually cool over time. This opens up an opportunity to study the formation and early evolution of planets (e.g. Keppler et al. 2018), which is difficult with methods such as RV due to the high activity levels of young stars. Direct imaging is the most versatile of all techniques, opening up opportunities for (e.g.) orbital monitoring, spectral characterization, variability characterization, and instantaneous study of multiple planets within the same data set. As we will detail in the next section, it is also the key technique for studying Earth-like planets in the habitable zones (HZs) of Sun-like stars.

## Toward habitable exoplanets

The transit and RV techniques are now at the level where Earth-sized planets (or even smaller, in the case of transits) can be detected in some circumstances. This is mostly in the cases of small orbits around small stars, since both of those aspects are favoured in RV and transit searches. Coincidentally, small stars also have their HZs close in, due to their lower bolometric luminosities. This means that the lower the mass of the star, the easier it is to detect planets in the classical HZ around it with RV/transits, provided that it remains bright enough for the instruments that observe it to acquire a sufficient signal. Indeed, some of the lowest-mass stars in the Solar neighbourhood now have detected planets (or planet candidates) in their HZs (e.g. Anglada-Escude et al. 2016; Gillon et al. 2017). In the future, there may be a promising option to characterize such planets with transmission spectroscopy, in order to map their atmospheric content and search for biomarker molecules.

However, this pathway toward studying habitability has several limitations. Firstly, it's not clear if planets around low-mass stars can be habitable in the first place. The HZs around mid/late M-type stars experience extreme levels of X-ray and UV radiation that dissociate water molecules, and strong stellar winds that may strip away the hydrogen component on rapid timescales. If the planet cannot hold on to its water, its equilibrium temperature (which defines the HZ) is inconsequential. In addition, flare intensities are much higher, which might erode the atmosphere; planets in the HZ are tidally locked, which impacts atmospheric dynamics and the ability to generate planetary magnetic fields; and low-mass stars have a very extended pre-main sequence phase, during which temperatures in what will eventually become the HZ are much too high for liquid water to occur. Secondly, even for the lowest-mass stars, transmission spectroscopy is very challenging, because it only probes the terminator region of a planet, which is a tiny fraction of its total area. Indeed, even the future ESA M-class mission Ariel, which is specifically designed for transmission spectroscopy, will not probe HZ planets, except for very particular purposes (e.g., probing whether the TRAPPIST planets have extended hydrogen atmospheres). Thirdly, and related to the previous points, transmission spectroscopy scales poorly asymptotically toward more ambitious missions, and especially in terms of reaching HZ planets around Sun-like stars, which is orders of magnitude beyond the difficulty of probing lower-mass stars with this technique.

In terms of detecting truly Earth-equivalent planets (same size, stellar host type, etc. as the Earth), all of the basic methods described above can contribute in different ways: Transits can provide a statistical census of Earth-like HZ planets through ESA's future M-class PLATO mission (Rauer et al. 2016). RV instruments are being developed with the aim of reaching sensitivity to Earth-equivalent planets, at least around quiet stars (Pasquini et al. 2010). *Gaia* is not sensitive enough to detect terrestrial planets, but astrometry concepts such as THEIA (PI: C. Boehm; see e.g. Janson et al. 2018 for details) exist that would be able to yield this sensitivity, and could therefore provide the invaluable resource of a complete census of Earth-like (and other) planets in the Solar neighbourhood, with mass measurements in each case. Direct imaging, as we will see below, can also detect Earth-equivalent planets with conceptual missions such as HabEx or LUVOIR.

However, the situation for exoplanet *characterization* is different: In this context, direct imaging is the only method that can feasibly characterize the atmospheres of Earth-like planets in the HZs of Sun-like stars, which in turn are the only kinds of planets that we already know can be habitable. This is a fundamental reason for why direct imaging is a key technology in exoplanet research, and the aim of most large-scale exoplanet-related mission concepts among the main (multi)-national space agencies.

## Direct imaging concepts

At present date, most high-contrast imaging is performed with ground-based telescopes. This is primarily due to availability of large aperture telescopes, which is a fundamentally important issue in exoplanet imaging due to the intrinsic faintness of planets in combination with the need to separate them spatially from the PSF of the parent star. So-called Extreme-AO (ExAO) systems enable the wavefront distortions imposed by the atmosphere to be corrected to a good extent, reaching Strehl ratios in excess of 90% in the near-infrared (where 100% would imply a perfectly diffraction-limited PSF), and coronagraphy helps in substantially reducing stellar light without affecting planetary light. Nonetheless, the fact that the coronagraph needs to operate on a non-perfect wavefront currently sets the limit for the achievable contrast from the ground, which is of order $10^{-7}$ at small separations with the help of differential imaging techniques in the best cases. For comparison, the final contrast required to image an Earth-like planet around a Sun-like star is $10^{-10}$ (in reflected light).

This fact has driven the development of space telescope concepts for exoplanet imaging, which avoids the issues related to the Earth's atmosphere. Such concepts are typically either based on internal coronagraphs or occulters, as described in the next section. The Nancy Grace Roman Space Telescope (previously known as WFIRST) is planned to have a sophisticated coronagraphic mode and will be of great help for developing such techniques, but its Hubble-class aperture (2.4m) is too small to allow for reaching true Earth analogs. Thus, plans are in motion for larger aperture space missions to accomplish this, including concepts such as HabEx and LUVOIR. We will not discuss these particular concepts in detail in this paper, but instead concentrate on general principles surrounding them. We note that a dedicated White Paper on these concepts was submitted by Ignas Snellen and coauthors and is included in this issue.

Another path in the context of direct imaging in space is the usage of nulling interferometry for thermal imaging in the infrared, with a formation-flying satellite fleet. We will not discuss this concept thoroughly either, but refer to the White Paper by Sascha P. Quanz and co-authors (in this issue) for more details.

## Occulters versus coronagraphs

Coronagraphs come in many different forms based on a range of different techniques, but the common feature in the context of this discussion is that they are internal to the telescope design, downstream from the primary mirror. This means that their contrast performance is necessarily diffraction related, with an inner working angle (IWA) determined by the telescope size (and observing wavelength), and it also means that the upstream optics needs to be extremely precise in order to maintain a perfect wavefront for the coronagraph to operate on. By contrast, an occulter is external to the telescope, and placed at a large separation away from it. The occulter edge is apodized in order to minimize diffraction around its edges. With ideal apodization and a sufficiently large size, an occulter can produce a $10^{-10}$ reduction of incoming starlight in the shadow behind it, and a telescope placed in the center of this shadow at a sufficiently large separation will be able to see companions to the star outside of the edges of the occulter, with the starlight already removed in the wavefront that hits the primary mirror. In this framework, the IWA is set by the occulter size relative to its distance from the telescope, independently of the telescope size. Furthermore, the specifications for the telescope optics become no more demanding than in regular non-coronagraphic imaging, so basically any telescope could be used or relatively easily adapted to operate in conjunction with the occulter.

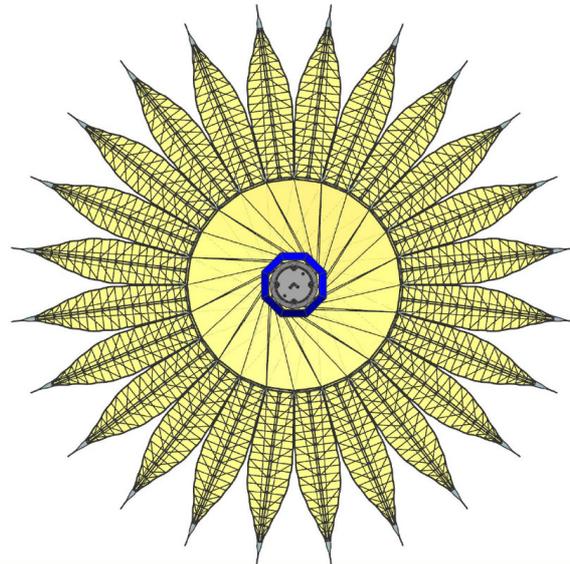

*Fig. 1: Occulter design example. The occulter consists of a central opaque circular structure, surrounded by a ring of petals that gradually narrow down outwards. This pattern helps to mitigate diffraction around the occulter edges to the extent that a very high contrast depth is acquired at the center of the shadow created by the occulter. Image credit: NASA*

The two techniques have their respective pros and cons. Operationally, coronagraphs are preferable over occulters, because it is vastly easier and faster to turn your telescope from the direction of your current target to the direction of the next one than it is to move the occulter thousands of kilometers from one target to the next. In general, the fact that the occulter requires a separate satellite unit obviously adds cost and complexity relative to a coronagraphic mission that (hypothetically) fulfils the same science requirements. However, a great advantage of the occulter is that it is simply technologically easier to acquire a high contrast than it is with a coronagraph. A full occulter-based mission with sensitivity to a large number of Earth-like HZ planets could be launched with minimal development given the current technological readiness – the financial aspect of such a mission is the only limitation to its feasibility.

## Occulter to Earth

So far, we have been discussing HabEx-like concepts where a space-based telescope operates in conjunction with a space-based occulter. In this setup, the telescope itself obviously contributes substantially to the mission cost even for a small aperture size, and this cost increases dramatically as the aperture size increases. Meanwhile, telescopes exist and are being constructed on Earth that are far larger than anything that exists or is planned in space. If an occulter could fit a telescope on the ground in its shadow with respect to a target star for a sufficiently long time (say, an hour), this would not only save the whole cost of launching the telescope, but it would also yield data of much higher signal-to-noise ratio (S/N) than would be achievable with the space-based unit.

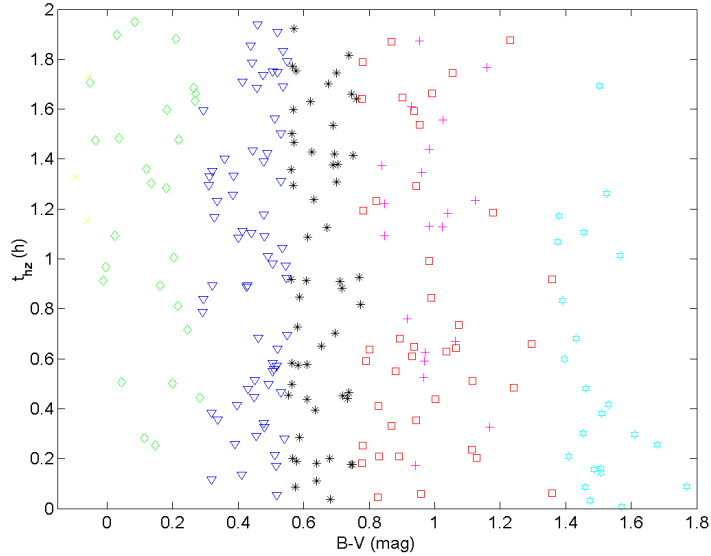

*Fig. 2: Stars for which Earths in the HZ (if present) would be detectable in less than 2 hours with an 8m-class telescope for an Earth-trailing occulter concept (Janson 2007). Possible targets would include everything from BA-type (yellow, green) to M-type (cyan) stars and even post-MS giants (magenta), but in particular a large number of FGK-type (blue, black, red) MS stars. © The Astronomical Society of the Pacific. Reproduced by permission of IOP Publishing. All rights reserved.*

We have previously seen that the atmosphere ruins the opportunity to acquire very high contrast imaging from the ground with coronagraphy, so one might intuitively expect that the same would be true for the occulter option. However, this is not the case. In the setup we are describing, the occulter operates in space, with no atmosphere around. It thus creates a wavefront with a high-quality shadow, which, for a telescope inside of the shadow, will be mapped as an image of the system with the central star already diminished by a factor $10^{10}$. The occulted image will be smeared by atmospheric seeing, but this is inconsequential with regards to the contrast, since the seeing halo of the star will anyway be $10^{-10}$ of its original brightness. It is however necessary to apply adaptive optics in order to collect light into the PSF core of any planets surrounding the star, since they would otherwise drown in the background emission from zodiacal and exozodiacal dust.

The key aspect of transmitting a shadow from space to the ground is to identify an orbit for the occulter in which the shadow can be maintained fixed with respect to a given coordinate on the ground for a long enough time to acquire a good S/N, at a reasonable fuel consumption. Meanwhile, the Earth is rotating and an object in orbit undergoes constant motion, which is the central challenge that needs to be overcome. Ideally, the orbit of the occulter during an observation should be such that the projected velocity of a telescope on Earth as seen from a target star is perfectly matched for an extended amount of time. While this can't be done perfectly due to the fact that the projected motion of a point on the ground changes continuously with the rotational phase of the Earth, small active boosts can be imposed on the occulter to create a good matching speed at a modest cost in fuel. For a single target, this is a simple problem to solve, but if we want to observe several targets, we also need a means to transition from one target to the next, and to time each new target up in phase such that the telescope and occulter line up with the target star simultaneously for each observation. This is a significantly more difficult problem.

In Janson (2007), we demonstrated a type of inclined Earth-trailing orbit that allows for shadow control on top of a ground-based telescope for timescales of order hours per telescope, as well as timed target transitions and continuous orbit corrections at a distinctly modest fuel demand. A central trade-off that allows this to work is that the sky coverage is limited to a band between declinations of approximately +/-5 deg on the sky. Recently, an alternative set of orbits has been developed by John Mather and collaborators at GSFC and JPL based on highly elliptical orbits[1]. This class of orbits is more directable, but also imposes a challenge for large transitions across the sky. These exercises demonstrate that a range of orbits does exist that can fulfil all the basic constraints for the occulter shadow control, with different trade-offs with regards to sky coverage and target sample size. In the end, the specific science requirements of the mission would guide the exact choice of orbits for the occulter.

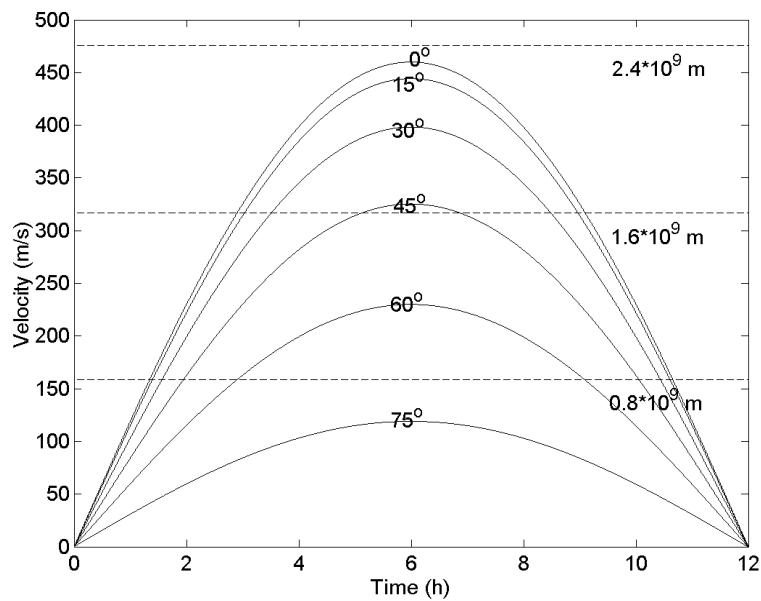

*Fig. 3: Matching projected speeds between ground-based telescopes at various latitudes (solid lines) and occulters at different separations (dashed lines). The occulter drifts in a controlled manner from one target to the next, and then boosts actively for a few hours during a target rendezvous in order to keep in exact pace with the Earth's rotation. From Janson (2007). © The Astronomical Society of the Pacific. Reproduced by permission of IOP Publishing. All rights reserved.*

---

[1] https://exoplanets.nasa.gov/internal_resources/1318/

## Occulter to the E-ELT

Transmitting an occulter shadow to the ground opens up the opportunity for aperture sizes much larger than what would be possible to launch into space. In particular, the ~40 m E-ELT[2] telescope is now under construction, providing the largest aperture available anywhere for optical/near-infrared purposes. Therefore, now is a good time to examine the prospects of combining a space-based occulter with the E-ELT, potentially providing extremely high S/N for Earth-analog planets relative to what is accessible with any other facility or combination of facilities. Indeed, the main purpose of this White Paper is to motivate technical studies for this purpose.

The contrast and IWA provided by a given occulter is intrinsically independent of telescope properties. The only additional constraint imposed by usage of the E-ELT relative to any other telescope is that the linear dimensions of the shadow necessarily have to be large enough to fit the entire aperture. An apodized occulter consists of a central opaque circle of diameter $D$, with narrowing petals extending out to approximately $2D$. The shadow dimensions cannot be larger than $D$, and will generally be a bit smaller for the part of the shadow that reaches $10^{-10}$ contrast, but can be quite close to $D$ (see e.g. Cash 2006). The exact size of this part of the shadow depends on the specific occulter design and becomes part of an optimization scheme when designing an occulter to provide a given contrast for a shadow of a given size over a given wavelength range. Here we simply note that the ~40 m diameter of E-ELT sets the constraint that the occulter needs to be at least 80 m from tip to tip of its petals. Such a size is well compatible with occulter-to-Earth concepts: The separation between the occulter and telescope in such concepts separately sets a requirement of occulter sizes of order 100 m in order to reach a sufficient contrast. Thanks to the large separation, the IWA in these concepts typically reaches a range of ~20-60 mas, i.e. sufficient to detect Earth analogs (1 AU) out to ~16-50 pc. This in turn gives access to Sun-like stars in the hundreds; how many of them can be observed is a question of fuel budget, as discussed more below.

---

[2] https://www.eso.org/sci/facilities/eelt/

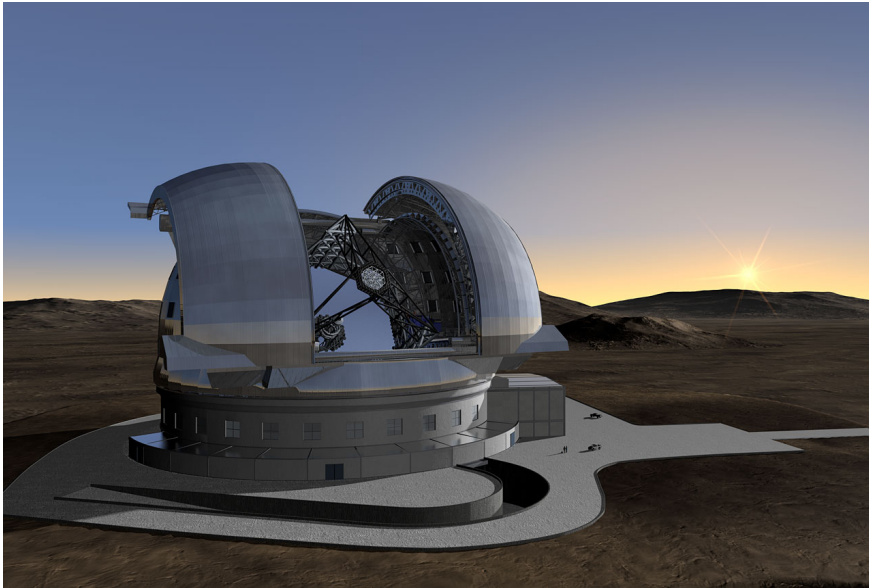

*Fig. 4: Rendering of the E-ELT telescope currently under construction at Cerro Armazones in Chile. Image credit: ESO*

## Technical developments

The main purpose of this document is to motivate studies regarding a number of technological developments that should be performed in order to further enhance the readiness of such a mission and evaluate key characteristics of the concept. We list the most important aspects below.

### Occulter design

Studies regarding occulter design and construction have largely been performed in the US so far, but it would be beneficial for the European community for such studies to be performed also in a European framework. The occulter has some necessary electronic components (such as boosters, power system, possibly a laser for guide star purposes) but is otherwise a potentially fairly simple mechanical system. Challenges include the issue of getting such a large structure into orbit (the American designs include intricate folding techniques), and ensuring a minimal degree of scattered or reflected sunlight, which is an important consideration primarily at the edges of the occulter. A key issue is to find ways of minimizing mass in the design. The mass directly sets the scope of the mission: Orbital calculations can be used to calculate delta-V values and to find the solutions that minimize the required delta-V for shadow control (which governs the available amount of observing time for each target) and target-to-target motion (which governs the total amount of targets that can be observed). Given an ideal orbit, the mass of the occulter is then the fundamental parameter that completely governs the total number of targets and the total hours of integration time available for the mission.

An interesting detail in this context is that NASA is considering a system (IRMA)[3] that could assemble structures in orbit, thus allowing for physically larger structures than can be launched with existing launcher facilities. It would not be able to assemble, e.g., full telescope units, due to the precision requirements of optical systems. However, a large 2D mechanical system such as an occulter may be an excellent match to such a facility. We also note that there is a potentially fruitful synergy with solar sails that could be utilized, since there are many overlapping criteria between what constitutes a good occulter and a good solar sail, in terms of large rigid 2D-structures with a high size-to-mass ratio.

**Large 2D structures in space**

As stated previously, the number of targets that can be observed in an Occulter-Earth mission is set by the fuel budget, which in turn is fundamentally governed by the dry mass of the occulter. In the Mather et al. concept discussed above, they use an expanded version of the design that has been developed for mission concepts such as HabEx. This yields an occulter dry mass of 14,000 kg, which for a wet mass of 22,000 kg allows for ~12 targets to be observed. The Janson (2007) orbital concept implies similar delta-V constrains, leading to similar target numbers if the same occulter solution would be used. However, from a European perspective, it is worthwhile to consider whether occulters could be constructed in Europe, and if so, if they could be made more lightweight. From first order principles, this seems feasible: there are concepts for solar sails of similar dimensions with estimated mass surface densities of 5.9 g/m$^2$ (Hughes et al. 2006, JSR 43, 4). The theoretical limit is even much lower: sails envisioned for Breakthrough Starshot would have mass surface densities of potentially below $10^{-3}$ g/m$^2$ (e.g. Heller & Hippke 2017). A circular occulter of radius 50 m at the more conventional value of 5.9 g/m$^2$ would have a mass of 46 kg. Even considering several layers (for protection against micrometeorites) and adding a support structure, an engine, etc., there appears to be room for substantial mass decrement relative to 14,000 kg, at least in principle. The design constraints for occulters and solar sails are obviously not completely identical, but there exist substantial similarities in the demand for large deployable 2D structures. A study on occulter design from a European engineering perspective and in communication with the solar sail community would thus be highly valuable, not only with regards to our highlighted concept, but indeed for any occulter mission in the future.

**Potential stepping stones for occulter missions**

In the context of any occulter mission, it is interesting to ask what the least costly scientifically meaningful mission would be, since this could act as a pathfinder mission to develop occulter technology in space. We argue that the answer is a single-boost eccentric orbit occulter mission to alpha Cen A and B, in conjunction with VLT/MUSE. The orbit would allow for ~1h of MUSE observation per star. This concept would require two occulters (one for A and one for B), but they can be much smaller (~28m tip-to-tip diameter) and less massive in total than the ELT options discussed above. With a contrast of $3\times10^{-10}$ and IWA of 160 mas, the mission would allow for detection and characterization of Earth analogs even at unfavourable orbital phases. MUSE-NFM spectra (R~3000) binned by a factor 30 yields S/N ~10 in 1h over much of the visible band (see Fig. 5). This is enough to clearly detect both $O_2$ at 760 nm and $H_2O$ around 900nm. We emphasize that the telescope component in this concept builds entirely on instrumentation that already exists, and whose performance is already well demonstrated. Such a mission could make use of existing occulter designs, and operational costs would be minimal compared to typical space missions since the whole mission phase would only last for a few hours (the occulters are launched into a highly elliptical orbit, observations commence at apogee, the occulters re-enter the atmosphere at perigee and are destroyed). No modifications are needed at the telescope level and the whole data infrastructure is already in place. The cost would therefore probably be strongly dominated by the launch cost. A risk element of such a mission is that it is unknown whether planets exist around alpha Cen A or B. Cheap mission concepts for figuring this out beforehand exist (e.g. Janson et al. 2018). We note that a single-boost occulter mission to Proxima Cen would be less favourable as a stepping-stone mission, since a much larger occulter would be required to fulfil the contrast and IWA requirements of Proxima Cen b.

---

[3] https://www.nasa.gov/sites/default/files/atoms/files/nac_tkortes_irma_nov2016_tagged.pdf

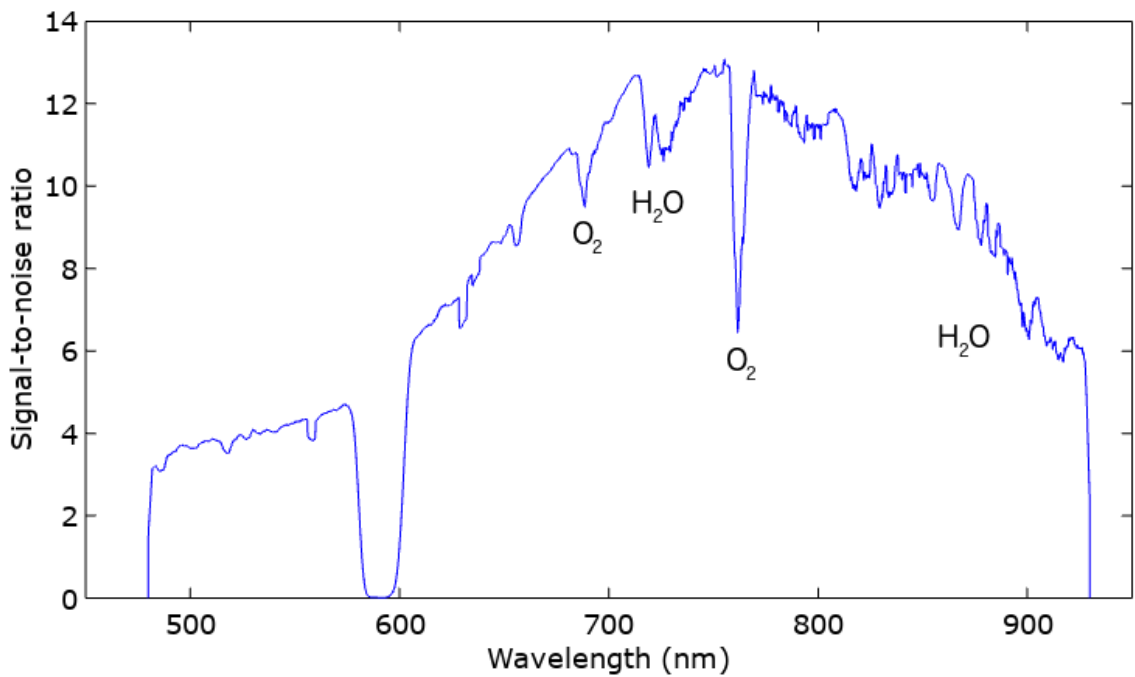

*Fig. 5: VLT-MUSE (factor 30 spectral binning) S/N as a function of wavelength for an Earth analog in the HZ of alpha Cen A, in 1h of integration. The $O_2$ line at 760 nm and the water band around 900 nm would be clearly distinguishable.*

**Optimal orbits**

As we have already touched upon, the orbital characteristics are crucial to the mission design. What constitutes an "optimal" orbit in this regard is context dependent. For example, if the scientific goal is to observe a limited number of very high-merit targets that are spread widely across the sky, elliptical orbits may be preferable, while if the goal is to observe a larger number of targets without prior information about relative merits, an Earth-trailing orbit may be a better option. There may also be orbital solutions for a given set of requirements that have yet to be discovered. Since there is an interplay between the science requirements and the orbit optimization, there needs to exist a dialogue between researchers and engineers on the orbital dynamics side. ESA would be the natural contact point within European astronomy for this purpose.

**E-ELT instrumentation**

It is obviously required that the E-ELT has suitable instrumentation for the observations that need to be executed once the occulter shadow lands on it. Since these requirements overlap in many ways with general requirements for ground-based high-contrast imaging, this should be relatively straightforward to implement, but it needs to be part of the plan that all of the requirements are definitely in place by the time the occulter launches, which in turn requires interplay between ESA and ESO in the mission planning phases. The main requirements for a suitable E-ELT instrument are as follows:

Firstly, it needs to operate in the visible wavelength range. Much of the high-contrast imaging that is performed from the ground is done at near-infrared wavelengths, but this is unsuitable for an occulter. The contrast that can be achieved with an occulter is wavelength dependent and becomes increasingly challenging at longer wavelengths. The exact wavelength range that is covered at sufficient contrast depends on the specific occulter design (e.g. Vanderbei et al. 2007), but in general terms, achieving good contrast at >1 micron wavelengths is excessively challenging with occulter technology.

Secondly, it needs to acquire as high Strehl ratio as possible. The Strehl ratio governs how much of the planetary signal is collected into its diffraction-limited PSF core, and how much spills out into its seeing-limited halo. A high Strehl ratio implies a high fraction of the light collected into the core. This fraction of the light is the only light that will be usefully observable to the telescope, because the halo light will blend together with the zodiacal and exozodiacal light pollution and thus not contribute beneficially to the S/N. The diffraction-limited core for a ~40 m telescope, on the other hand, is fully distinct from the diffuse background, to the extent that the observation becomes photon noise limited. This second requirement is partly at tension with the first one, in that it is more difficult to acquire a high Strehl ratio in the visible than in the near-infrared. However, the existing ZIMPOL (Schmid et al. 2018) arm of the SPHERE instrument at VLT demonstrates that high Strehl ratios are in fact possible in the visible, routinely providing Strehl ratios of 50% and above in the red part of the spectrum.

Additionally, it is preferable if the instrument has an Integral Field Unit (IFU) capability. With an IFU, a spectrum for all point sources in the field of view can be acquired upon first visit, without prior knowledge of their locations. This is very useful for an occulter concept, where re-visiting a given target can be quite challenging. For example, the Earth-trailing orbit concept described above can only re-visit targets on a yearly basis, which is good for common proper motion conformation and orbital characterization since a lot of motion has occurred in between; but for the exact same reason, it is bad for spectroscopic follow-up with, e.g., a slit, since it cannot be predicted exactly where the companion will re-occur. If a priori orbital information exists, for example from RV or astrometry data, this is less of an issue.

An instrument that could potentially deliver on all of these points is PCS[4], formerly EPICS, which is foreseen as a second-generation instrument on the E-ELT with first light in the 2030s (PI: M. Kasper, ESO). In designing an occulter mission, it should be carefully studied whether PCS is suitable for delivering the necessary data and if any modifications can be done to better accommodate this purpose, or whether even a special-purpose instrument may be a desirable option.

## Summary

In this paper, we have outlined some of the main aspects of an occulter-to-Earth concept, where an occulter in space projects a deep shadow onto a specific location on Earth – specifically, the site of E-ELT – and some of the main technical considerations that we argue should take place within the European community, with collaborations around the world whenever applicable. A key goal in exoplanet science, as well as in astronomical science as a whole, is to evaluate prevalence and conditions for habitability and life in the Universe. This can only be accomplished by detecting and characterizing a statistical sample of potentially habitable worlds. The concept presented here constitutes one promising way of accomplishing this within sensible budgetary constraints. Other possible concepts for doing so exist, and are presented in White Papers submitted by Snellen et al. and by Quanz et al. The existence of several different concepts for similar scientific purposes should not be seen as a sign of divisiveness within the exoplanet community. Rather, it displays that there are several promising paths toward a very broadly commonly accepted scientific goal. Depending on scientific and budgetary constraints, it may benefit the European community to choose one path or another upon careful evaluation. The occulter-to-Earth concept presented here represents a path in which Europe could potentially play a leading role in the detection and characterization of habitable planets beyond the Solar System. The E-ELT will be a thoroughly unique facility, unmatched in size and scope by any other existing or planned telescope. By combining it with an occulter in space, it would become a formidable tool for observing the habitable Universe.

---

[4] http://ao4elt3.arcetri.astro.it/proceedings/info_12804.html

**The necessity of direct imaging in space**

A clear picture arising from the Voyage 2050 workshop (held in Madrid, Spain, 29-31 October 2019), both from the talks in the exoplanet session as well as other White Papers, is the central importance of space-based direct imaging for making progress toward characterization of potentially habitable exoplanets. This is not a new insight, but has been noted before in several past roadmaps and long-term strategic documents among the large (multi)-national space agencies[5]. However, so far there has been limited implementation of this scientific priority in actual space missions (with exceptions including the high-contrast arm of the Roman Space Telescope concept). The direct imaging concepts that have been presented in the different Voyage 2050 White Papers reflect a variety of possible paths toward characterizing Earth-analog planets, spanning a range of techniques, administrative constellations, and cost caps. It is beyond the scope of this paper to attempt to assess their relative merits. We simply note that, regardless of which particular path is chosen, the most important aspect in the outcome of the scientific evaluation of Voyage 2050 is that it needs to reflect the overall importance of direct imaging in the quest for habitable worlds, in order to ensure that this priority carries through into ESA's actual mission selection.

---

[5] For example: http://www.nap.edu/catalog.php?record_id=12951

[6] M. Janson, 2007: *Celestial Exoplanet Survey Occulter: A Concept for Direct Imaging of Extrasolar Earth-like Planets from the Ground*, The Publications of the